\documentclass[twoside,11pt]{article}
\usepackage{amsmath}
\usepackage{amssymb}

\usepackage{blindtext}

\usepackage{jmlr2e}
\usepackage[
top    = 0.55in,
bottom = 0.8in,
left   = 0.55in,
right  = 0.55in]{geometry}
\usepackage{wrapfig}
\usepackage{float}

\usepackage{lastpage}

\firstpageno{1}

\begin{document}

\begin{center}
    \Large \bf Clustering Inductive Biases with Unrolled Networks
\end{center}

\begin{center}
    \textit{Jonathan Huml\textsuperscript{1}, Abiy Tasissa\textsuperscript{2}, Demba Ba\textsuperscript{1}}\\
    (1) Harvard John A. Paulson School of Engineering and Applied Sciences, (2) Tufts University Department of Mathematics
\end{center}

\textbf{Summary.} The classical sparse coding (SC) model represents visual stimuli as a linear combination of a handful of learned basis functions that are Gabor-like when trained on natural image data. However, the Gabor-like filters learned by classical sparse coding far overpredict well-tuned simple cell receptive field profiles observed empirically. While neurons fire sparsely, neuronal populations are also organized in physical space by their sensitivity to certain features. In V1, this organization is a smooth progression of orientations along the cortical sheet. A number of subsequent models have either discarded the sparse dictionary learning framework entirely or whose updates have yet to take advantage of the surge in unrolled, neural dictionary learning architectures. A key missing theme of these updates is a stronger notion of \emph{structured sparsity}. We propose an autoencoder architecture (WLSC) whose latent representations are implicitly, locally organized for spectral clustering through a Laplacian quadratic form of a bipartite graph, which generates a diverse set of artificial receptive fields that match primate data in V1 as faithfully as recent contrastive frameworks like Local Low Dimensionality, or LLD \citep{lld} that discard sparse dictionary learning. By unifying sparse and smooth coding in models of the early visual cortex through our autoencoder, we also show that our regularization can be interpreted as early-stage specialization of receptive fields to certain classes of stimuli; that is, we induce a weak clustering bias for later stages of cortex where functional and spatial segregation (i.e. topography) are known to occur. The results show an imperative for \emph{spatial regularization} of both the receptive fields and firing rates to begin to describe feature disentanglement in V1 and beyond.\\

\textbf{Background. } Let $\mathbf{Y} = [\mathbf{y}_1, \ldots, \mathbf{y}_n] \in \mathbb{R}^{d \times n}$ be a set of $n$ visual stimuli each of dimension $d$. For unknown dictionary $\mathbf{A} = [\mathbf{a}_1, \ldots, \mathbf{a}_m] \in \mathbb{R}^{d \times m}$ and latent representation $\mathbf{X} \in \mathbb{R}^{m \times n}$, typically with $m << n$, classical sparse coding attempts to represent $\mathbf{Y} = \mathbf{AX}$ using a sparse prior over $\mathbf{X}$ via an $\ell_1$ norm (columnwise) to enforce sparsity and learn both $\mathbf{A}$ and $\mathbf{X}$ via alternating gradient descent. However, while neurons fire sparsely, the receptive fields in V1 also vary smoothly over the cortical sheet, which motivates our Laplacian smoothness term. To construct this term, we build a bipartite graph Laplacian with edge $x_{ij} \in \mathbf{X}$ between stimuli $\mathbf{y}_i$ and basis function $\mathbf{a}_j$, with no self-connections within the stimuli or basis function vertices. We denote the coordinate representation of these vertices by $\mathbf{R} = [\mathbf{Y} \quad \mathbf{A}] \in \mathbb{R}^{d \times (n+m)}$ and the weight matrix $\mathbf{W} \in \mathbb{R}^{(n+m) \times (n+m)}$ built for this bipartite graph by:
\begin{align*}
    \mathbf{W} = \begin{bmatrix} \mathbf{0}_{n \times n} & \mathbf{X}^T \\ \mathbf{X} & \mathbf{0}_{m \times m} \end{bmatrix}
\end{align*}
The graph structure implies a constraint $x_{ij} \geq 0$ and $\mathbf{X}^T \mathbf{1} = \mathbf{1}$, or that each column of $\mathbf{X}$ lives on the probability simplex. The graph Laplacian is then defined as $\mathbf{L} = \mathbf{D} - \mathbf{W}$, where $\mathbf{D}$ is a diagonal matrix formed by summing over the rows of $\mathbf{W}$. The quadratic form of this graph Laplacian is given by the trace of $\mathbf{RLR}^T$, which is subject to the original convexity constraint $\mathbf{Y} = \mathbf{AX}$. This program enforces smoothness as the graph Laplacian is a discretization of the divergence of the gradient along the stimuli manifold. However, where does sparsity enter the model? By rewriting the quadratic form, we can observe a conceptual similarity to the classical $\ell_1$ penalty:
\begin{align}
\text{trace}(\mathbf{RLR}^T) &= \text{trace} (\mathbf{Y}^T \mathbf{YI}) + \text{trace} (\mathbf{A}^T \mathbf{A} \text{diag}(\mathbf{X1})) - 2 \text{trace} (\mathbf{R}^T \mathbf{RW})\\
&= \sum_{i,j} x_{ij} ||\mathbf{y}_i - \mathbf{a}_j||_2^2 \qquad \text{s.t.} \qquad \mathbf{Y} = \mathbf{AX}
\end{align}
The probability simplex constraint on $x_{ij}$ conceptually acts as an $\ell_1$ prior as in classical sparse coding, but weighted by the distance between a given stimuli and the set of the receptive fields (hence the name weighted $\ell_1$ penalty for this quadratic form). If this distance is high, this will penalize the firing rate in the minimization procedure. Thus, from the unification of sparsity and smoothness in the early cortex emerges a type of simple, early-stage specialization. Experimentally, we therefore wish to verify that the receptive field profiles learned by the objective function (Eq. 2) are ``brain-like,'' and that the firing rates maintain a clustering structure. We unroll \citep{unrolling} this objective into a recurrent autoencoder architecture (WLSC) which implicitly solves the weighted $\ell_1$ problem via backpropagation.

\begin{figure}[H]
\centering
\label{fig:first-img}
    \includegraphics[scale=0.6]{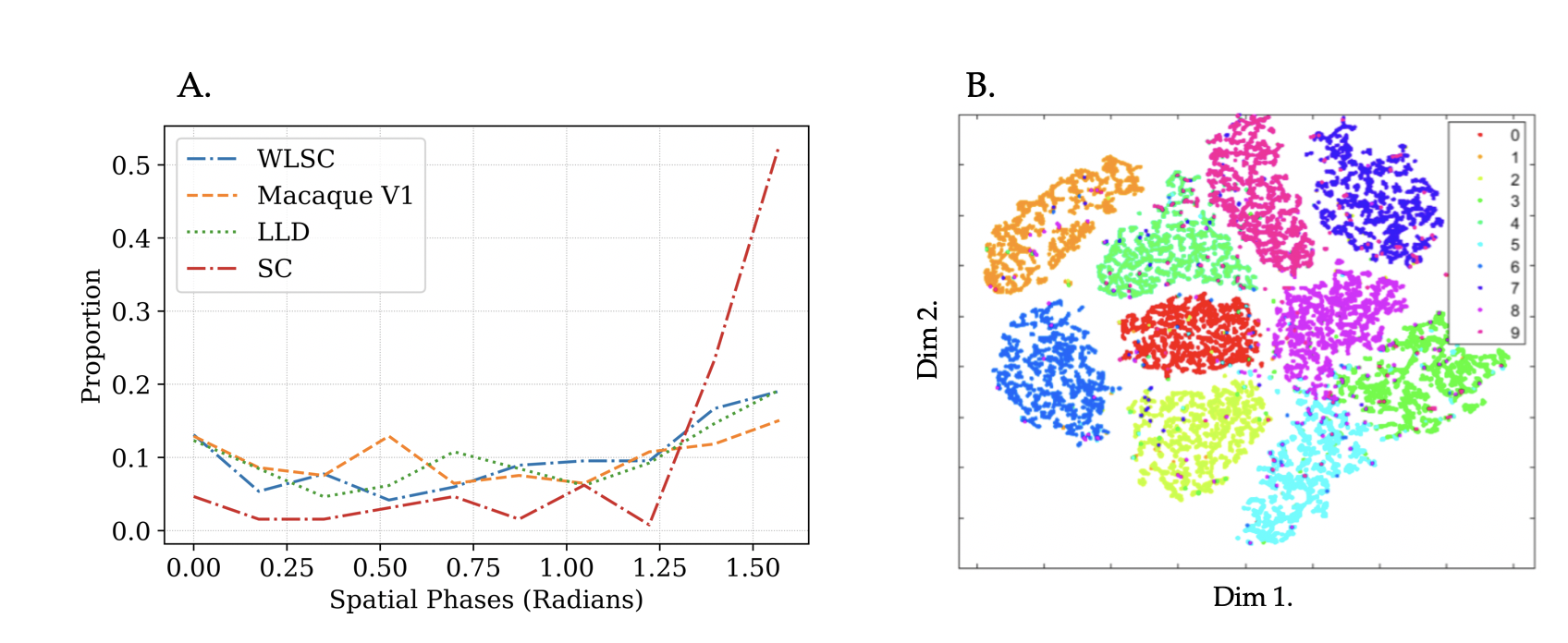}
    \caption{(A) The proportion of simple cell receptive field Gabor spatial phases learned on natural images for the weighted-$\ell_1$ penalty (WLSC) match physiological data despite being conceptually similar to sparse coding objectives (B) Latent representations learned by the autoencoder on MNIST, when projected to 2D for visualization by t-SNE, maintain a natural clustering structure}
\end{figure}

\textbf{Results.} While the weighted $\ell_1$ penalty uses core ideas of the original sparse coding framework, the spatial phase distribution shown in Figure 1(A) is far more faithful to physiological data \citep{ringach}, matching recent contrastive frameworks like LLD, which posit that reconstruction is not a desirable computational assumption. When trained on natural image patches, we observe a greater number of broadly-tuned cells in the weighted $\ell_1$ model compared to the original sparse coding model. Moreover, we see that smooth and sparse codes imply a clustering structure as in Figure 1(B), where we project the learned latent representations on MNIST into a 2D space for visualization purposes (assuming that separability in a lower dimensional space implies separability in a higher dimensional space).   This computational result is of particular interest for the construction of topographic artificial neural networks (ANNs), which in practice have typically used features from more generic architectures and subsequently imposed specific topographic constraints in models of inferotemporal cortex where specific modules (faces, objects, etc.) are known to exist. However, can this topographic structure be derived from more generic constraints given these early-stage receptive fields that are arranged in a spatially plausible fashion? The results presented here, and interpretations thereof, raise questions regarding the extent of feature-sharing between various topographic modules. 



\vskip 0.2in
\bibliography{sample}

\end{document}